# Spoofing Against Spoofing: Towards Caller ID Verification In Heterogeneous Telecommunication Systems

SHEN WANG, University of Warwick, United Kingdom
MAHSHID DELAVAR, University of Warwick, United Kingdom
MUHAMMAD AJMAL AZAD, School of Computing and Digital Technology, Birmingham City University, United Kingdom
FARSHAD NABIZADEH, Nar Co., Iran
STEVE SMITH, trueCall Ltd, United Kingdom
FENG HAO, University of Warwick, United Kingdom

Caller ID spoofing is a global industry problem and often acts as a critical enabler for telephone fraud. To address this problem, the Federal Communications Commission (FCC) has mandated telecom providers in the US to implement STIR/SHAKEN, an industry-driven solution based on digital signatures. STIR/SHAKEN relies on a public key infrastructure (PKI) to manage digital certificates, but scaling up this PKI for the global telecom industry is extremely difficult, if not impossible. Furthermore, it only works with IP-based systems (e.g., SIP), leaving the traditional non-IP systems (e.g., SS7) unprotected. So far the alternatives to the STIR/SHAKEN have not been sufficiently studied. In this paper, we propose a PKI-free solution, called Caller ID Verification (CIV). CIV authenticates the caller ID based on a challenge-response process instead of digital signatures, hence requiring no PKI. It supports both IP and non-IP systems. Perhaps counter-intuitively, we show that number spoofing can be leveraged, in conjunction with Dual-Tone Multi-Frequency (DTMF), to efficiently implement the challenge-response process, i.e., using spoofing to fight against spoofing. We implement CIV for VoIP, cellular, and landline phones across heterogeneous networks (SS7/SIP) by only updating the software on the user's phone. This is the first caller ID authentication solution with working prototypes for all three types of telephone systems in the current telecom architecture. Finally, we show how the implementation of CIV can be optimized by integrating it into telecom clouds as a service, which users may subscribe to.

Category: • **Security and privacy → Authentication**.

## 1 INTRODUCTION

Telephone scams have been increasing at an alarming rate, especially during the Covid pandemic [14]. According to a survey in 2022, 1 in 3 Americans (33%) report having been targeted by phone scams, causing a total loss of nearly $40 billion in the past 12 months [26]. In these scams, fraudsters frequently use *caller ID spoofing* to modify the caller's phone number in order to hide identity or to pretend to call from trusted sources (e.g., banks, tax revenue offices) [32].

Caller ID spoofing has been known possible since the calling line identification (CLI) was first introduced in 1987 as a telephone service to display the incoming call number [9]. Sometimes, there are legitimate reasons for the caller to modify the number, e.g., showing a single outgoing number for an organization or a toll-free number for the callee to dial back. In traditional phone systems, modifying a caller ID requires special hardware or access to the telecom infrastructure. However, with Voice over IP (VoIP) [11], it has become effortless to modify the caller ID using the

---

software. Once a modified number is permitted by the originating network, it will be trusted by the subsequent networks without validation. The ease of modifying caller ID has unfortunately enabled many frauds [6].

Besides inflicting financial losses on victims, caller ID spoofing attacks also reduce public trust in the telephone infrastructure. Ofcom, the telecom regulator in the UK, has been warning the public not to trust the caller ID display, and instead, users should "hang up and call the phone number to check whether the call was genuine" [14]. The Federal Communications Commission (FCC) has similar advice: "Don't answer calls from unknown numbers. If you answer such a call, hang up immediately." [6]. While these warnings serve to raise public awareness of the untrustworthiness of the caller ID display, they also have a significant side effect: according to YouGov, nearly 90% of people simply stop answering phone calls with unknown numbers [31]. Legitimate personal and business calls are blocked as well.

To restore the public confidence in caller ID display and stop the spoofing attacks, the FCC has recently mandated telecom providers in the US to implement STIR/SHAKEN by 30 June 2022 (gateway providers by 30 June 2023) [7]. STIR/SHAKEN represents a suite of protocols developed by an Internet Engineering Task Force (IETF) working group to combat caller ID spoofing. It works by attaching a digital signature as part of a Session Initiation Protocol (SIP) header together with a VoIP call. In practice, the digital signature needs to be accompanied by a certificate chain in the transmission to allow verification by the receiving party.

The FCC describes STIR/SHAKEN as "an industry-standard caller ID authentication technology" [7]. But actually, STIR/SHAKEN does not authenticate any caller ID. Instead, it authenticates the *originating carrier* where the call is made (or the *gateway carrier* for international calls that arrive inbound at the gateway), based on the carrier's exclusive possession of a private signing key. Arguably this solves a different problem (which may not exist as any major issue in the telecom industry). The problem of authenticating caller IDs remains unsolved and is left to carriers: carriers must attest, as part of the digital signature, whether the caller is authorized to use the phone number. The difficulty here is to distinguish between legitimate and illegitimate modifications of the number. Unfortunately, carriers do not always have the knowledge to tell them apart (if they do, caller ID spoofing would have been a much easier problem to tackle). For example, when a user modifies the VoIP caller ID to a mobile phone number, the carrier may not know if the user is authorized to use the number that belongs to a different carrier. To cater for this, STIR/SHAKEN introduces A, B, and C levels of attestations (for full, partial, and gateway attestations respectively) to indicate the carrier's knowledge with different levels of confidence under different conditions. Level A attests that the caller is authenticated and is authorized to use the number; level B attests that the caller is authenticated but it is unknown if they are authorized to use the number; level C attests that the call is signed at a gateway without knowing if the caller is authenticated or is allowed to use the number. Interpreting validation results for signed calls of different levels has proved hard for users [18].

Apart from the ambiguity in the definition of "authentication", STIR/SHAKEN suffers from two inherent limitations in the system design. First, it critically relies on trusted certificate authorities (CAs) to certify signing keys in a public-key infrastructure (PKI). VoIP networks normally involve a PKI when using SSL/TLS to protect the data transmission in certain paths, but the PKI we discuss here in STIR/SHAKEN is a *new* infrastructure. To spur the deployment of STIR/SHAKEN, the FCC has appointed several telecom companies in the US as the CAs. To comply with STIR/SHAKEN under the FCC rule, telecom providers in the US shall pay these CAs subscription fees for the issuance of certificates, normally based on the company's annual revenue (subject to a minimum fee) [22]. Although the FCC has been urging a global adoption of STIR/SHAKEN, it is extremely unlikely that the FCC-appointed CAs will be trusted by all other countries. (Similarly, if China appoints its own CAs, it is equally unlikely that they will be trusted by the FCC.)

Second, STIR/SHAKEN involves the transmission of digital signatures and a chain of certificates (several kilobytes) as part of the signaling data. The original design is to support only IP-based systems (e.g., SIP), which leaves the traditional non-IP systems (e.g., SS7) unprotected [18]. Although there are retrospective proposals to support STIR/SHAKEN in SS7 systems, e.g., by transmitting signature data out-of-band through a trusted third party, such a trusted third party is difficult to find in reality. The FCC acknowledges that "the STIR/SHAKEN framework is only operational on IP networks", but requires that "providers using older forms of network technology to either upgrade their networks to IP or actively work to develop a caller ID authentication solution that is operational on non-IP networks" [7]. However, the "caller ID authentication solution" for non-IP networks has not been specified, which leaves a gap in the regulatory rules [32].

Public data show that since STIR/SHAKEN was mandated in June 2022, this solution has not been as effective as expected. First of all, after the mandate of STIR/SHAKEN, the number of robocalls actually went up and reached a record of 5.5 billion calls in October 2022 [21]. Many of the robocalls are now signed with STIR/SHAKEN to look more legitimate. Among all the signed calls with the B-attestation, about a quarter are robocalls; for calls signed with the C-attestation, about a third are robocalls [23]. Many of these signed robocalls present a different caller ID from where they are calling. The statistics [20] also show that although nearly 70% of the outbound VoIP calls are signed with STIR/SHAKEN, only 15-24% of the calls received by the terminating networks have valid signatures; for many calls, the signatures are removed as they traverse intermediate non-IP networks. It is also reported that many calls are signed with wrong attestation levels, which should not be surprising given that STIR/SHAKEN only authenticates the "carrier" and the attestation of the caller ID is entirely based on the carrier's "word of mouth" [32].

STIR/SHAKEN represents an industry-driven approach: representatives from several telecom companies form the core of an IETF working group to specify a signature-based framework called STIR, followed by implementation details called SHAKEN [8]. Some of these companies were later appointed by the FCC as the CAs. Besides serving a trusted role, there is also an economic incentive to be a CA, since other companies are obliged to pay them subscription fees in compliance with the FCC regulation. So far, this industry-driven solution has received limited scrutiny from the security research community. In particular, alternatives to STIR/SHAKEN have not been sufficiently studied. Given the fundamental limitations of STIR/SHAKEN and the prevalence of caller ID spoofing attacks, we believe that it has become more urgent than ever to explore more secure and effective solutions to stop spoofing attacks. Since many spoofing calls originate from overseas [18], it is crucial that we address it as a *global* problem rather than any regional or country-specific problem. Furthermore, instead of treating caller ID authentication for IP and non-IP networks as separate problems, we propose to tackle them together with a unified solution.

In this paper, we propose Caller ID Verification (CIV). CIV authenticates caller ID based on a challenge-response protocol. As we will explain, it does not require a PKI and works with existing heterogeneous networks (SS7/SIP), hence addressing the two major limitations of STIR/SHAKEN. Our solution does not require any trusted third party and can be deployed by updating the software on the user's phone (or switches in the telecom cloud). This follows a bottom-up approach as opposed to STIR/SHAKEN's top-down approach. There are previous bottom-up proposals [12, 4], which *probabilistically* infer the authenticity of the caller ID based on heuristics. By contrast, CIV *deterministically* authenticates the caller ID based on a challenge-response protocol (detailed comparisons with the related work in Section 7). Our contributions are summarized below.

- We propose CIV, a new bottom-up solution to authenticate the caller ID based on a challenge-response process without depending on a PKI.
- We present concrete prototypes of CIV for landline, cellular and VoIP phones and show how they work across heterogeneous telecom networks (SS7/SIP).
- We systematically evaluate the performance of CIV in PSTN, cellular, and SIP networks to show the feasibility. This is the first demonstration of a caller ID authentication solution that works on all three types of phone systems across heterogeneous networks.

## 2   BACKGROUND

Over time, telecommunication systems have evolved to be exceedingly complex, spanning heterogeneous networks. In this section, we explain the relevant background below.

### 2.1   Heterogeneous networks

**Public Switched Telephone Network (PSTN).** PSTN represents circuit-switched telephone networks, which are traditionally connected by copper wires to transmit analogue voice signals. Later, Integrated Services Digital Network (ISDN) was developed to allow the digital transmission of data over copper lines, which also introduced new telephony features, such as voice mail, call forwarding, and caller ID/name display. The Signaling System 7 (SS7) is the dominant protocol to control calls in PSTN, including call setup, connection, tear-down, and billing [5]. maintaining dedicated wires for PSTN has become increasingly expensive. BT has announced a plan to phase out PSTN by 2025 [19]. However, for the foreseeable future, PSTN will still be used in many parts of the world [28].

**Cellular Network.** Cellular technology allows transmitting voice data wirelessly rather than over wired connections. The first generation of mobile phones (1G) used analogue signals, and the handset simply sent the serial number in the air, which was vulnerable to cloning attacks [1]. The second generation (2G) adopted digital technology with encryption. GSM was introduced in 1992. Each GSM handset has an embedded SIM card that stores a unique *international mobile subscriber identification* (IMSI) number and a secret key for authentication and encryption. In 1993, cdmaOne was introduced as another 2G standard based on CDMA. The third generation (3G) entered service in 2003, providing a faster data rate by adopting a spread-spectrum technology. The fourth generation (4G) was rolled out in 2009, followed by the fifth generation (5G) in 2019. While SS7 has been used as a core signaling protocol in 2G and 3G, it has been replaced by SIP in 4G and 5G. For interoperability, 4G and 5G still need to interconnect with previous-generation networks before they are completely phased out. For this reason, SS7 is still supported by nearly all wireless carriers today [2].

**Voice-over-IP (VoIP).** In VoIP, voice data are digitized, compressed, and routed over an IP network [1]. Currently, the dominant signaling protocol in VoIP is the Session Initiation Protocol (SIP). SIP borrows many of its syntax and semantics from HTTP (hypertext transfer protocol), but it also inherits many weaknesses of HTTP [30]. For example, the SIP header is unprotected, which makes it trivial to spoof a caller's identity. A primary benefit of VoIP is that the running cost is low as no dedicated wires are needed. Unlimited local and long-distance calls are often included in a bundle, which greatly saves costs for users, but at the same time also enables spammers and scammers to do robocalling (with a spoofed caller ID) at little cost [27]. The interconnection between VoIP and other networks (PSTN and cellular) is achieved through gateways, which are responsible for the conversion of different signaling types.

## 2.2 Caller ID spoofing

There are two pieces of information to identify a caller: a caller ID (phone number) and an optional caller name. A spoofing attack involves modifying the phone number, name, or both. Calling line identification (CLI) was first introduced in 1987 in the US as a telephone service to allow displaying of the caller's number on the receiver's phone. Later, Caller Name (CNAM) was introduced as a separate service to allow the calling party to specify a name associated with the caller ID. Typically, a caller name is limited to 15 characters, including alphanumerics, commas, and spaces. Special characters (e.g., &, @, etc) are not allowed for ordinary users but are permitted for business users. In the US, the originating carrier does not normally send the caller's name when initiating a call. Instead, the carrier registers the caller name in shared CNAM databases. It is the responsibility of the terminating carrier to look up the CNAM databases (paying a "dip" fee) based on the received number and deliver the retrieved caller name to the callee's phone. When access to CNAM databases is not available (e.g., outside the US), the caller name is sent directly to the called party together with the caller ID; in the case of VoIP, it is included in the SIP "From" header.

Modifying the caller ID/name is always possible in a telecommunication system. As part of CNAM, a user can freely modify the caller name (subject to basic checks such as the length). Modifying the caller ID is less straightforward. In PSTN, the telephone user is authenticated to the local switch through a dedicated wire. The caller ID is generated by the local switch, and an ordinary user cannot modify it. However, if the calling party is behind a Private Branch eXchange (PBX) which is connected to a local switch via Primary Rate Interface (PRI), PBX can modify the caller ID [3]. The modified caller ID is usually accepted by the switch without validation and passed on to the rest of the network. In a cellular network, the IMSI number stored in the SIM card is used by the switching center to identify the caller ID of a cellular user [17]. Similar to PSTN, an attacker has a limited chance to modify the caller ID. In VoIP, the caller ID, together with the optional caller name, is specified as part of the SIP "From" header. However, the header is unauthenticated. Hence, a user can arbitrarily change the caller ID and the caller name in the header. Once the modified header is permitted by the originating carrier, it will pass the subsequent networks without validation.

Carriers often allow users to use a different caller ID than what they are assigned with [12]. According to the Truth in Caller ID Act of 2009, modifying the caller ID is permitted, unless it is done "with the intent to defraud, cause harm, or wrongfully obtain something of value" [10]. There are legitimate reasons for modifying the default caller ID [12], e.g., to provide a toll-free number for the receiver to call back, or to display a central phone number for all outgoing calls from an organization. We do not regard these cases as spoofing attacks and will distinguish them from illegitimate spoofing based on whether the caller possesses the claimed phone number (details in Section 3).

## 2.3 Dual-tone multi-frequency

Dual-tone multi-frequency (DTMF) is one important element in our solution, so we explain it in more detail here. DTMF is a telecommunication signaling system that was first invented by the Bell System in 1963 [29], and then standardized and adopted globally. It uses a pair of the 8 predefined frequencies (hence called dual-tone) to encode one of the 16 phone button presses, including digits (0-9), *, #, A, B, C and D. DTMF is especially useful for enabling users to provide input of short digits over a phone line, e.g., to enter a PIN in telephone banking, or navigate menus in an interactive voice response (IVR) system.

DTMF signals can be transmitted through the voice channel as part of an audio stream (in-band) or a separate control path (out-of-band), depending on the underlying telecom system (see Figure 1).

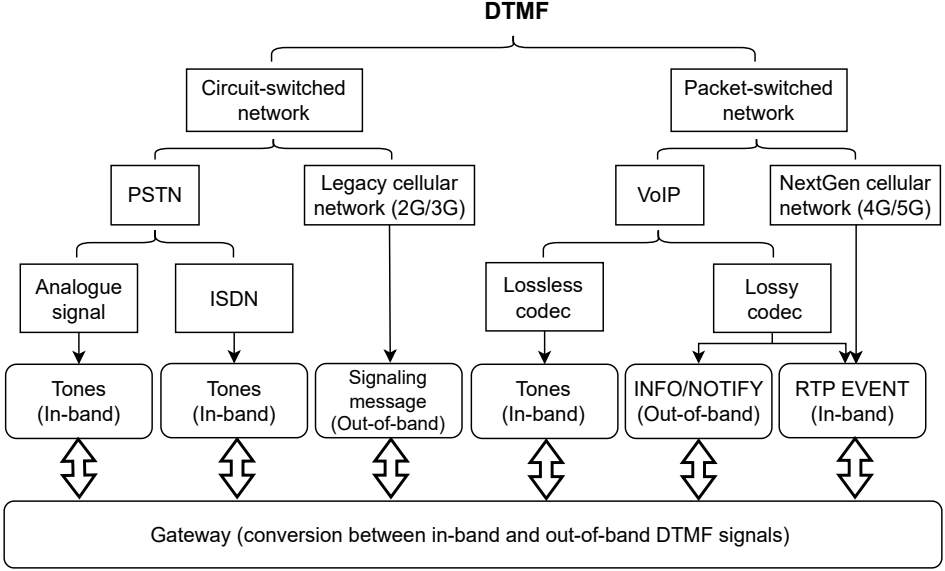

Fig. 1. Overview of DTMF transmission

Gateways are responsible for the seamless transmission of DTMF tones across different phone networks, handling the conversation between in-band and out-of-band signals wherever necessary.

**Circuit switched network.** In a PSTN network (analogue lines or ISDN), DTMF tones are typically transmitted as in-band signals in the range of human voice frequencies ($300 - 3400$ Hz). However, in a cellular network like GSM, the data transmission rate is significantly slower than in a wired network. To achieve a comparable voice quality in cellular networks, the voice data are heavily compressed before transmission. Widely used compression algorithms remove frequencies in the audio data that are insensitive to human ears so a phone conversation is not affected, but the loss of certain frequencies impacts the decoding of DTMF tones. To ensure reliable transmission of DTMF in a circuit-switched cellular network (2G/3G), DTMF data are transmitted *out-of-band* as signaling messages via a control channel separate from the voice communication [29].

**Packet switched network.** In a VoIP network, DTMF tones can be transmitted in-band together with the media stream if no compression or a lossless codec (e.g., G.711) is used. When a lossy codec is used (e.g., G.729), there are two ways to transmit DTMF: 1) in-band as part of the media stream but in a special Real-time Transport (RTP) Event packet based on RFC 4733; 2) out-of-band in a SIP INFO (RFC 6086) or NOTIFY (RFC 3265) message. In 4G/5G networks, DTMF tones are transmitted as RTP Event packets according to RFC 4733. In all cases, only the digital values of DTMF (not the analogue tones) are transmitted through the networks. When needed, the analogue sound of the DTMF tones is regenerated and played locally on the phone to inform users about sending/receiving DTMF.

## 3 OUR PROPOSED SYSTEM

Figure 2 shows an overview of our solution in heterogeneous networks. Our proof-of-concept implementations only require updating the software on the user's phone; implementation details for landline (in conjunction with a trueCall box), cellular and VoIP phones will be explained in

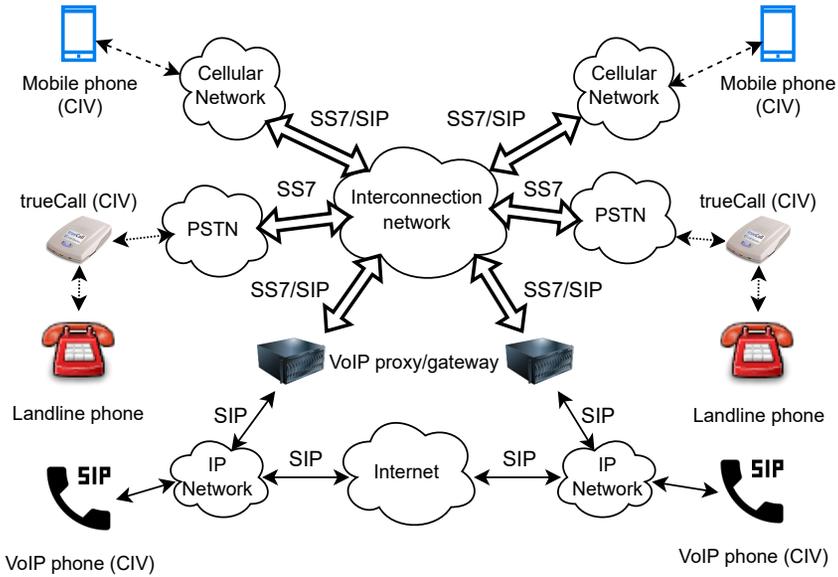

Fig. 2. CIV for heterogeneous telecommunication networks

Section 4. In Section 6.4, we will explain how to optimize the performance of CIV by integrating it into the networks.

### 3.1 Threat model

In our threat model, we assume that the attacker is able to arbitrarily modify the caller ID/name when initiating a call. The modified caller ID and name are permitted by the originating carrier and pass through subsequent networks. However, we assume the attacker is not able to intercept calls in the telecommunication system. We note that a powerful adversary can intercept calls through the Law Enforcement Monitoring Facility (LEMF), SIM swap, and SS7 hacking [1], but this is beyond the capability of ordinary telephone scammers behind the number spoofing attacks.

### 3.2 Protocol description

We name the caller 'Alice', the callee 'Bob', and the spoofing attacker 'Eve'. Eve tries to impersonate Alice by spoofing Alice's caller ID as his outbound number. Here, we focus on the spoofing of the caller ID (number), not the caller name, since the latter can be addressed by caller ID filtering [24] or reverse number look-up [25].

In CIV, the authentication of the caller ID is based on the possession of the claimed phone number. The intuition follows from how people verify the caller ID of an incoming call in real life by manually calling back the number. As an example, suppose that Bob receives a call displaying his bank's telephone number, which matches exactly the number shown on the back of Bob's bank card. But the number is actually spoofed[1]. Ofcom advises that Bob should hang up and call back the bank's contact number [14]. The manual call-back verification is slow, and tedious and may incur a charge for Bob. The goal of CIV is to turn this manual verification into an automated one (see Fig. 3) with minimum delay and cost.

---

[1]In many countries, the phone numbers printed on the back of bank cards are for *inbound* calls only: customers use them to contact banks but banks *never* use them to call customers. Ofcom includes these numbers in a "do not originate" (DNO) list and requests telecom operators to block them at the network level for *outbound* calls. However, complying with DNO is not (yet) a legal requirement. Blocking these calls by operators is not guaranteed.

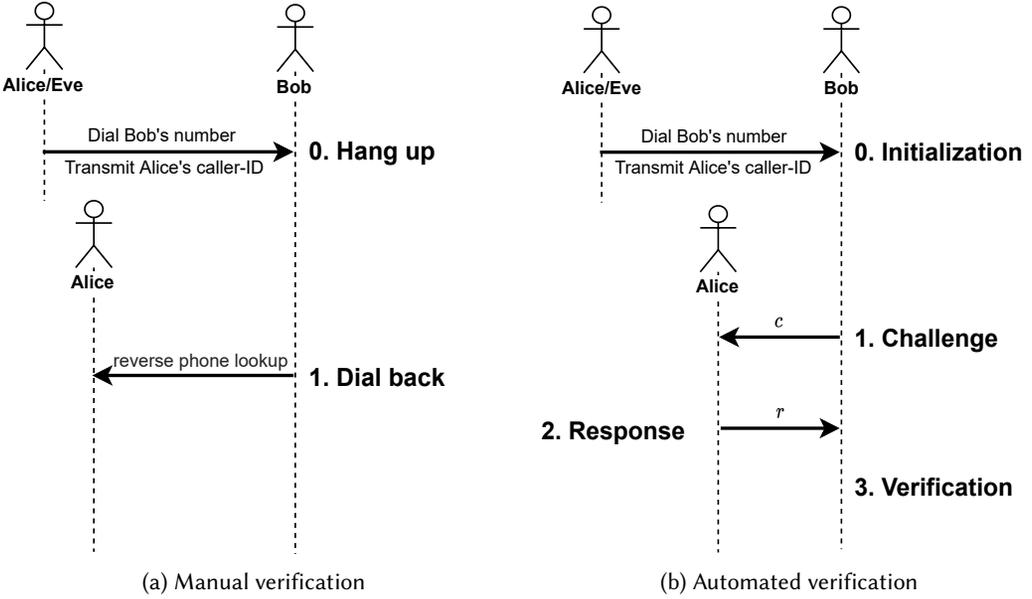

Fig. 3. Intuition behind CIV

In CIV, we assume that Alice (caller) actively wants to have her caller ID verified. The rationale is that she wants Bob to see the *verified* status of her caller ID so that Bob is more likely to answer the call. The cost to Alice is a possibly longer duration of a call, needed for carrying out the verification process. We assume that Alice is willing to pay for this cost. Here, the cost and Alice's incentive are aligned. In the initial call flow, Alice dials Bob's number and transmits her caller ID/name. In this flow, Alice indicates support for CIV, e.g., by adding a flag in the caller name. In practice, this flag may be a special character or a string of characters added during the registration of the caller name in CNAM databases (in our SIP-based prototype of CIV, we indicate support for CIV by transmitting a caller name appended with '*'; for other prototypes, we assume that a flag has been added to CNAM).

Upon receiving Alice's call, CIV on Bob's phone holds the incoming call and performs the following challenge and response protocol to verify the caller ID.

(1) **Challenge** – Bob's CIV calls back the number on the caller ID display and transmits a random $n$-digit number as a challenge $c$. In our design, we choose $n = 4$ (see Section 5 for evaluation of other values). If Alice is a genuine caller, she will receive $c$.

(2) **Response** – To prove the possession of the purported caller ID, Alice's CIV simply sends the same $n$-digit number as the response $r$ to Bob on a separate channel.

(3) **Verification** – After receiving $r$, Bob's CIV checks if $r \overset{?}{=} c$. If they are equal, the caller ID is "verified"; otherwise, it is "not verified".

After the above challenge-response process, Bob's phone starts ringing, displaying the caller ID together with the verification status. Figure 4 shows an illustration of the CIV user interface on a mobile device. For a landline phone without a display, we play an audio message to inform the user if the caller ID has been verified when they pick up the phone (we have implemented this by using a trueCall box which we will explain later). We emphasize that even if the caller ID verification fails, CIV still connects the call (Figure 4.b). In other words, CIV never blocks a call; it only adds additional information about the *verified* status of the incoming caller ID. The caller will likely experience a longer delay in connecting the call, but the delay is not perceivable by the callee. We have implemented this operation mode for all three types of phones (landline, cellular and VoIP). It is possible to implement an alternative operation mode, in which Bob's phone rings as soon as it receives a call and performs verification in parallel. This is to support an "emergency call" scenario (e.g., for 911) as described by Mustafa et al. [12]. While the implementation of this "emergency call" mode is possible in CIV, one might question if an emergency service such as 911 really needs this verification process since they already have privileged access to all the call detail records and can trace any incoming call (if needed). For this reason, we do not propose this emergency scenario as any main operation mode in CIV.

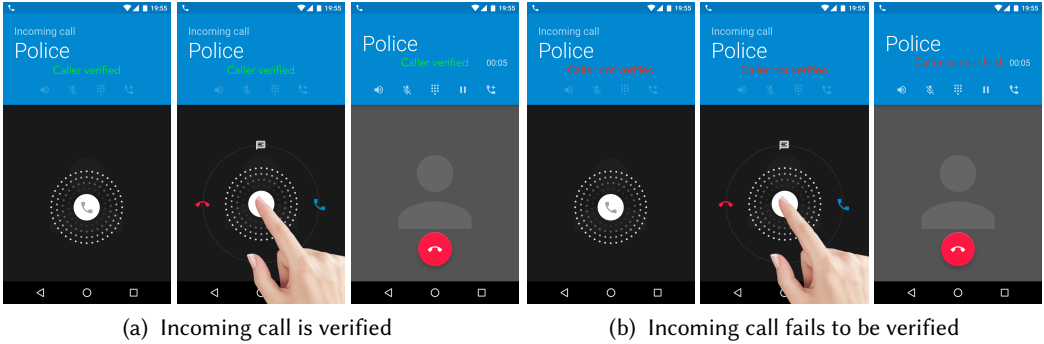

(a) Incoming call is verified                    (b) Incoming call fails to be verified

Fig. 4.  Illustration of the CIV user interface

Our protocol leverages the call-back session as a trusted channel to reach the owner of the purported caller ID (the attacker cannot intercept the call based on the threat model in Section 3.1). Note that CIV on the callee's phone only starts the verification process if the caller indicates support for CIV, since completing this process requires the cooperation of the caller. The user may further configure CIV on the phone to perform verification only when the displayed caller ID is a domestic number or a non-premium number.

The main challenge in realizing CIV is how to transmit the challenge and response in heterogeneous networks reliably, with minimum delay and cost. As explained in Section 2, caller ID spoofing has always been possible at a system level in all telecom networks. Here, we propose to leverage the facility of number spoofing to build a defensive mechanism to combat caller ID spoofing attacks, i.e., spoofing against spoofing. More concretely, Bob uses $c$ as his (spoofed) caller ID and makes an *abandoned* call to Alice. Here, there is no call charge to Bob. Alice's CIV receives a *missed* call and recognizes that it is a verification call (based on the format of the number but we can also make it explicit, e.g., by including the information in the accompanying caller name). The challenge $c$ is extracted from the caller ID. The missed call allows Bob to send a 4-digit message through telephony networks without cost. Alice may use the same spoofing method to send back the response, but we propose to transmit the response via the initial call using DTMF since it is much quicker (see Section 5 for experimental results).

### 3.3 Distinguishing legitimate/illegitimate spoofing

An effective caller ID authentication mechanism should be able to distinguish a legitimate modification of a caller ID from an illegitimate one (i.e., a spoofing attack) [12]. By design, STIR/SHAKEN does not make this distinction and relies on the carrier's 'word of mouth' in the attestation. CIV distinguishes these cases based on whether the caller possesses the phone number, hence being able to respond to a challenge sent to that number. As an example, a VoIP phone user wants to modify the caller ID to his mobile phone number. To support CIV, he simply needs to configure the mobile phone to forward the verification call containing the challenge to the VoIP phone. In another example, a caller is behind a PBX and wants to use a single outgoing phone number for the organization. In this case, PBX needs to keep the state of currently active outbound calls. It can assign a *random* index (say three digits for up to 999 simultaneous calls) for each call and include the index in the caller name of an outbound call so that when a verification call (containing the challenge and the index) arrives, it can forward the challenge to the corresponding caller. The CIV on the caller's phone will process the challenge automatically.

## 4 PROTOTYPES

In this section, we present proof-of-concept CIV prototypes for landline, cellular and VoIP phones across heterogeneous networks (SIP/SS7).

### 4.1 Overview

We have implemented prototypes of CIV for all three types of phones: landline, cellular and VoIP phones (see Figure 5). Our proof-of-concept prototypes are done under two levels of constraints. The first is on an infrastructural level: we have no cooperation from telecom providers. Therefore we can only update the software on the user's phone. The second is on a platform level: we are constrained to work with only the available APIs provided by different phone development platforms, as explained below.

(1) **VoIP platform**. We use the Ozeki VoIP Software Development Kit to develop a Windows-based SIP phone that implements CIV. The SIP phone works with a commercial third-party VoIP server, as well with our own VoIP servers, which we set up by using the open-source Asterisk software. Our VoIP servers are connected with the public SS7/SIP networks through SIP trunking.

(2) **Android platform**. We use Android phones (Nexus 5) and the third-party phone API available on Android (6.0.1) to build a phone app that implements CIV. The Android API supports the call waiting function (which allows placing a call on hold to engage in another call) but does not support transmitting DTMF in a call.

(3) **trueCall box**. We use a third-party nuisance call-blocking device, called trueCall [24], to implement CIV and connect the modified trueCall box to a landline phone. This allows us to control calls to a landline phone without having to modify its firmware. In contrast to Android, trueCall supports sending DTMF during a call but does not support the call-waiting function.

Figure 6 shows an overview of the CIV prototypes developed under various constraints. In general, there are two ways to transmit the challenge/response: using 1) spoofed CLI; 2) DTMF. The first method essentially uses CLI as a side channel but requires access to the facility of modifying the caller line identity: i.e., caller ID/name. This is only possible on our SIP platform. It is possible to modify the CLI for PSTN and cellular phones, but this needs to be done at the local switches or switching centers. The second method transmits data through DTMF. Without any local platform constraint (like on our SIP platform), this can be done efficiently by adding only one call setup.

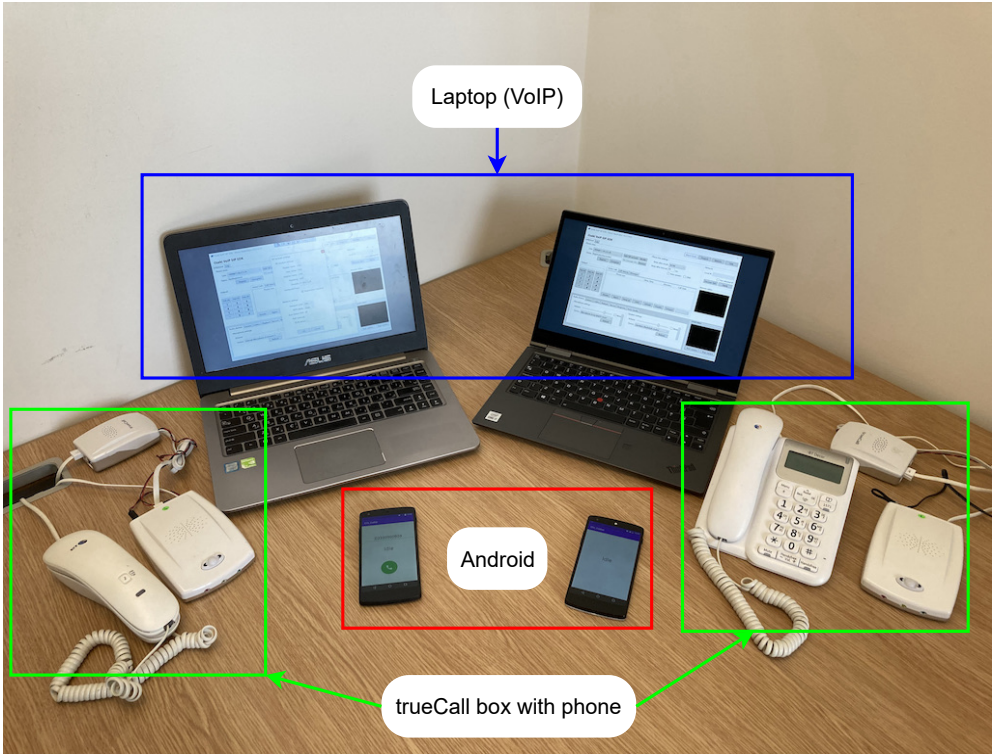

Fig. 5. Equipment setup

However, with the constraints of the trueCall/Android platforms, we need to add two call setups. This does not stop the proof-of-concept demonstration of CIV, but it increases the delay (see Section 5 for details).

## 4.2 CLI-based prototypes

These prototypes assume access to the facility of modifying the CLI (Case 1 in Figure 6). Based on our SIP platform, we have done two prototypes, 'CLI/CLI' and 'CLI/DTMF', which use two different methods to send 'challenge/response' respectively. ('DTMF/CLI' is another possibility but is not recommended as we explain below.)

Figure 7 summarizes the flows in the 'CLI/DTMF' implementation using SIP phones. In the initial call setup, Alice (caller) indicates support for CIV, e.g., by appending a flag in the caller's name. When CNAM databases are used for registering caller names, a special flag for a caller name indicates that an associated phone number is ready to be verified. In Step 1, when Bob receives an initial call from Alice, CIV on Bob's phone answers the call and puts it on hold. In Step 2, it then makes a verification call to the displayed incoming call number using a spoofed CLI (the challenge $c$). The verification call is immediately abandoned by Bob's CIV, and the initial call is taken off hold. The CIV on Alice's phone can distinguish it from ordinary missed calls and retrieve the challenge $c$ (random 4 digits that are received as the CLI). This process is handled transparently (users do not need to see the missed call). Finally, in Step 3, CIV on Alice's phone sends a response $r$ (same 4 digits) through DTMF via the initial channel established in Step 1. When receiving the response $r$, CIV on Bob's phone checks if it is equal to the challenge $c$. It concludes that the caller ID is verified if they are equal, and unverified otherwise. At this point, CIV starts ringing with the display of a caller ID along with the verification status.

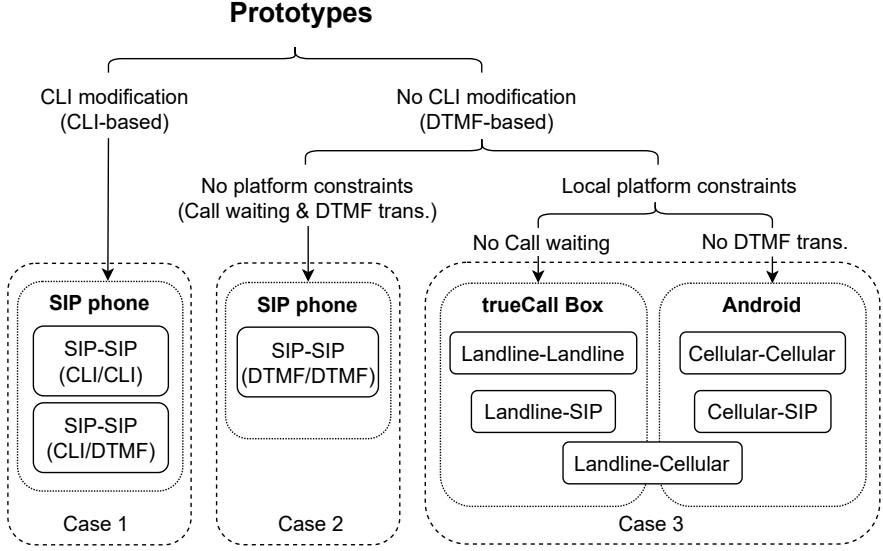

Fig. 6. Overview of CIV prototypes

An alternative implementation is to use 'CLI/CLI' to send 'challenge/response' respectively. The first two flows are the same as 'CLI/DTMF' in Figure 7. However, in Step 3, the response $r$ is sent to Bob as the modified CLI in an abandoned call, instead of using DTMF via the initial call. We have implemented this alternative 'CLI/CLI' approach, however, using 'CLI/CLI' incurs a longer delay than 'CLI/DTMF' because sending $r$ through CLI requires a full call setup (see experimental results in Section 5).

In theory, it is also possible to use 'DTMF/CLI' to send 'challenger/response' respectively. However, this option is not recommended due to the possible call charge to the callee (for sending the challenge through DTMF). Furthermore, it has no performance advantages over other options.

**Modifying CLI**. In our CLI-based prototypes, we need access to the facility of modifying CLI. On an Asterisk VoIP server, this can be easily done by invoking the `Set(CALLERID())` method to modify the default caller ID and name for an outgoing call in a configuration file. This is normally how the CLI spoofing is carried out from a VoIP platform. However, the modification is statically defined in a configuration file while we need to dynamically modify the CLI during a call. Also, for the proof-of-concept prototypes, we want to limit ourselves to modifying the client phone only (not the server). We use Ozeki VoIP SDK to build a client softphone to communicate with a VoIP server based on the standard SIP protocol. The Ozeki SDK does not support modifying the caller ID directly from the client, but it allows modifying the caller name by setting a new value for the `CallerDisplay` property. With access to this function, we set a special (*) flag in the caller name in the initial call to indicate support for CIV, and embed the challenge/response digits in the modified caller name.

**Transmitting DTMF**. Ozeki supports transmitting DTMF in-band as `RTP EVENTS`. We invoke the `StartDTMFSignal()` and `StopDTMFSignal()` functions to send DTMF tones as part of the media stream. In telecommunication terminology, the duration of a DTMF tone is called the 'mark' time while the gap between two consecutive DTMF tones is called the 'space' time. Telecommunication standards require the mark and space to be at least 40 ms (RFC 4733); in our SIP prototypes, we set both to be 50 ms. We note that the Ozeki SDK does not allow sending DTMF through an *out-of-band* channel from the client. However, if CIV is implemented on the SIP server, sending DTMF *out-of-band* will be possible, e.g., as an `INFO` or `NOTIFY` message.

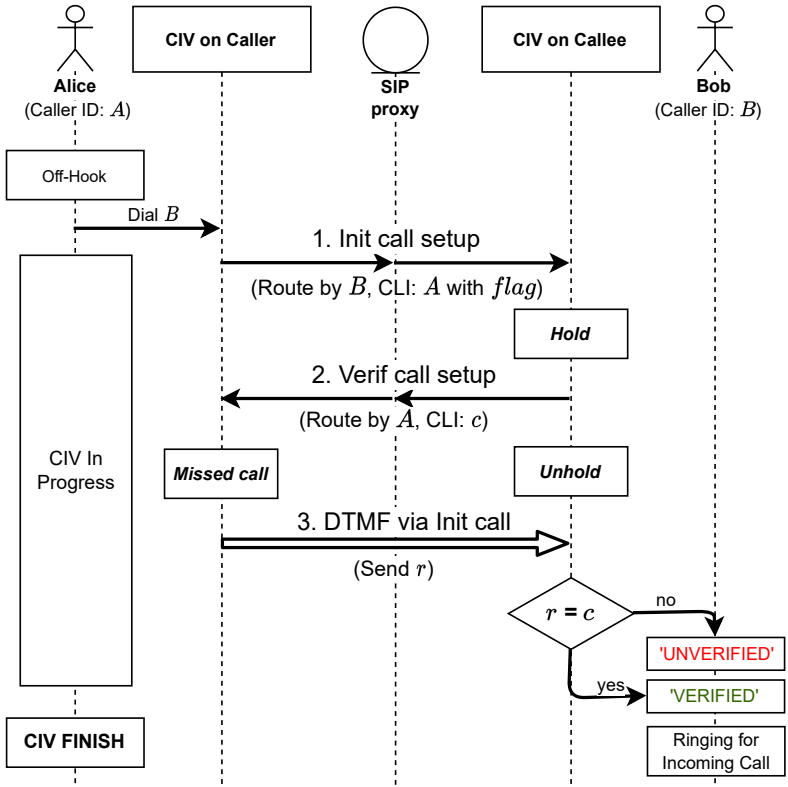

Fig. 7. CIV using CLI/DTMF to send challenge/response

### 4.3 DTMF-based prototypes

In our existing telephony systems, the end phones may not have access to the facility of modifying the CLI, or the local carrier may not permit such modification. In this case, it is still possible to implement CIV by using DTMF to send both the challenge and the response.

*4.3.1 No local platform constraints.* We assume that the user's phone supports the *call waiting* function (hence it can hold an incoming call) and is able to transmit DTMF during a call. We use SIP phones to implement a prototype for this case (i.e., Case 2 in Figure 6).

Figure 8 summarizes the implementation of this prototype which uses 'DTMF/DTMF' to send 'challenge/response' respectively. Upon receiving an initial call from Alice with a flag in the caller name, CIV on Bob's phone holds the call and meanwhile starts a *verification call* to send a challenge $c$ through DTMF. Once CIV on Alice's phone receives the challenge, it hangs up the verification call and sends a response $r$ using DTMF through the initial call. When CIV on Bob's phone receives $r$ and determines that it is equal to the challenge $c$, it confirms Alice's caller ID as *verified*; otherwise, it is *unverified*. Finally, Bob's phone starts ringing, with a display of the caller ID along with the verification status. This prototype is reasonably efficient as it needs only one additional call setup to the challenge (the response is sent through an existing initial call channel rather than a new call).

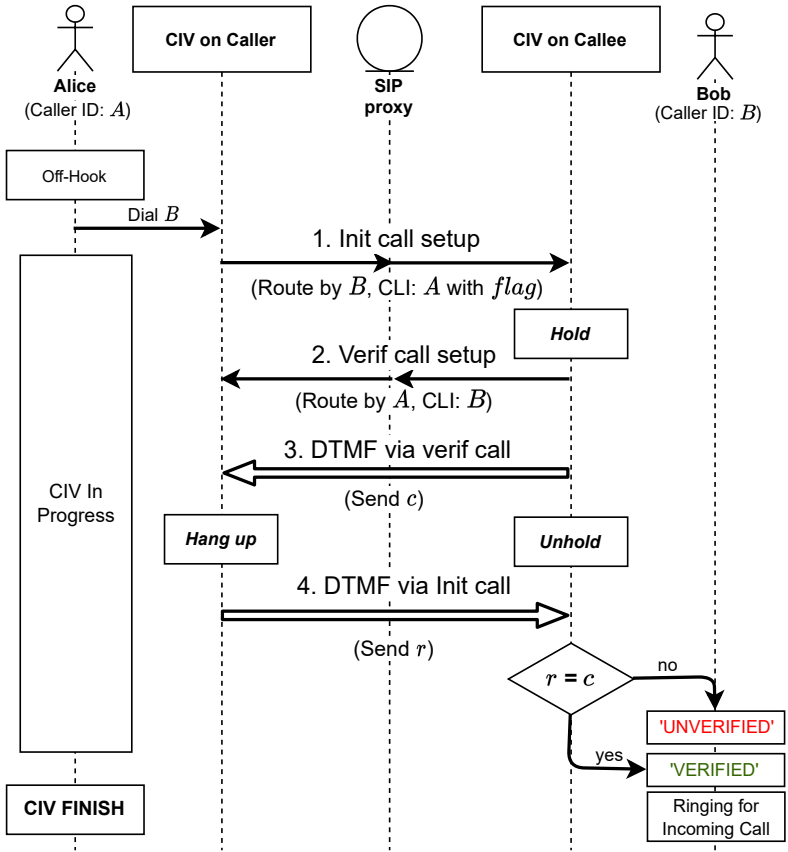

Fig. 8. Using DTMF/DTMF to send challenge/response

*4.3.2  With Local platform constraints.* For non-SIP phones (landline/cellular) used in our proto-typing, there are certain platform constraints (Case 3 in Figure 6). In particular, the trueCall box does not support the call waiting function (while it supports sending in-call DTMF). The Android third-party phone API does not support sending in-call DTMF (while it supports the call waiting function).

**trueCall**. We need a way to implement the CIV protocol on an analogue phone, however, we cannot modify its firmware. To address this issue, we use trueCall, which is a commercial call-blocking device, designed to protect elderly and vulnerable people from nuisance and scam calls. The hardware box contains a micro-controller that performs various call-control functions (e.g., off-hook, hang-up, ringing). We are able to modify the software in the trueCall box to implement CIV, but the hardware has its limitations – specifically, it does not support the call-waiting function. (Supporting the call waiting function in trueCall is possible but needs an extra chip in the hardware.)

In order to implement CIV on the trueCall hardware three call setups are needed. After Bob's CIV (implemented in trueCall) receives the initial call (Step 1), it cannot hold the call. Instead, it terminates the call and then starts a verification call (Step 2) to send the DTMF challenge. Alice's CIV will need to start a new call (Step 3) to send the DTMF response. Overall, three call setups are required. Since the challenge and response process is handled by CIV, it is transparent to users except that the caller will experience a longer wait for the extra call setup (details in Section 5).

**Android**. We use the Android third-party phone API (6.0.1) to build a CIV-enabled phone app for Android phones (Nexus 5). There are several implementation challenges that we need to overcome. First, the third-party API on Android only allows us to control one call at a time, but in CIV, we need to handle the verification call in parallel to the initial call. To overcome this limitation, we use Java reflection to invoke the hidden system service `TELEPHONY_SERVICE` to access the internal interface `ITelephony` in run-time. This allows us to hold the initial call using the system API while performing the verification call using the third-party API. Second, Android only allows a user to send DTMF manually by pressing keys during a call but does not support doing this programmatically. We sidestep this limitation by appending the DTMF digits as an extension after the phone number separated by a comma: e.g., '5555555555,1234' where ',' indicates a short pause (2 seconds on an Android phone). During dialing, Android first calls '5555555555', waits for 2 seconds, and sends '1234' through DTMF automatically. This method is commonly supported on mobile phones, allowing a user to directly reach an extension number behind a PBX without having to talk to an operator. We use the same method to send the challenge and response through DTMF, but this means that we need to specify the DTMF values upon dialing rather than during the call. As a result, we will need three call setups in the implementation. Finally, the third-party API does not support the automatic recognition of DTMF tones. To overcome this limitation, we set `AudioManager` to the `MODE_IN_CALL` mode, which allows the DTMF tones to be played via the speaker. Once CIV receives the DTMF tones through the microphone, it converts them into digits by analyzing the frequencies based on the Fast Fourier Transform (FFT). This allows us to build a proof-of-concept prototype of CIV on Android phones, but the speed of recognizing DTMF is significantly limited.

For CIV between two Android phones, we cannot send the response using DTMF via the initial call as previously done for SIP phones (see Figure 8). Instead, we use a new call to send the response (by appending the 4 digits to the phone number as an extension during dialing) through DTMF. This means we need three call setups as opposed to the 2 call setups. However, for a SIP phone calling an Android phone, the implementation is not affected by the Android's in-call DTMF limitation, and it can still be done in 2 call setups. We will present a detailed performance evaluation in Section 5.

## 5 EVALUATIONS

We evaluate the overhead of running CIV for calls between landline, cellular and VoIP phones. For each call scenario, we measure the delays of the CIV protocol during the day (10:00-13:00) and at night (21:00-24:00). We take 15 measurements at each time slot and present the average in Figure 9. The breakdowns of the measurements at night are presented in Figure 10 (the breakdowns during the day are basically the same). The breakdowns comprise four components: 1) setting up a verification call; 2) transmitting the challenge; 3) setting up a response call (if needed); and 4) transmitting the response.

### 5.1 SIP platforms

First, we evaluate the performance of CIV prototypes based on SIP platforms (see Case 1 and 2 in Figure 6). As shown in Figure 9, SIP 'CLI/DTMF' incurs the lowest 4.7 sec latency, followed by 5.7 sec in SIP 'DTMF/DTMF' and 6.8 sec in SIP 'CLI/CLI' during 21:00-24:00. The latency measurements during 10:00-13:00 are approximately the same. 'CLI/DTMF' requires one call setup to send the challenge through the (modified) CLI, and uses the existing initial call channel to send the response through DTMF. It is quicker than 'DTMF/DTMF' as it does not have the cost of transmitting the

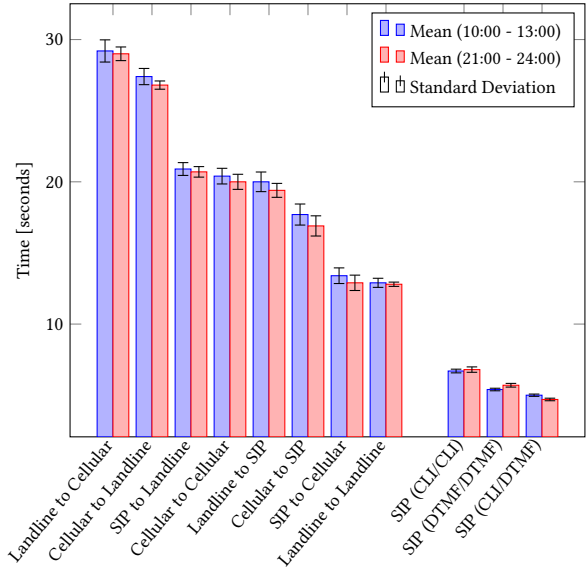

Fig. 9. Delays in CIV between different phones

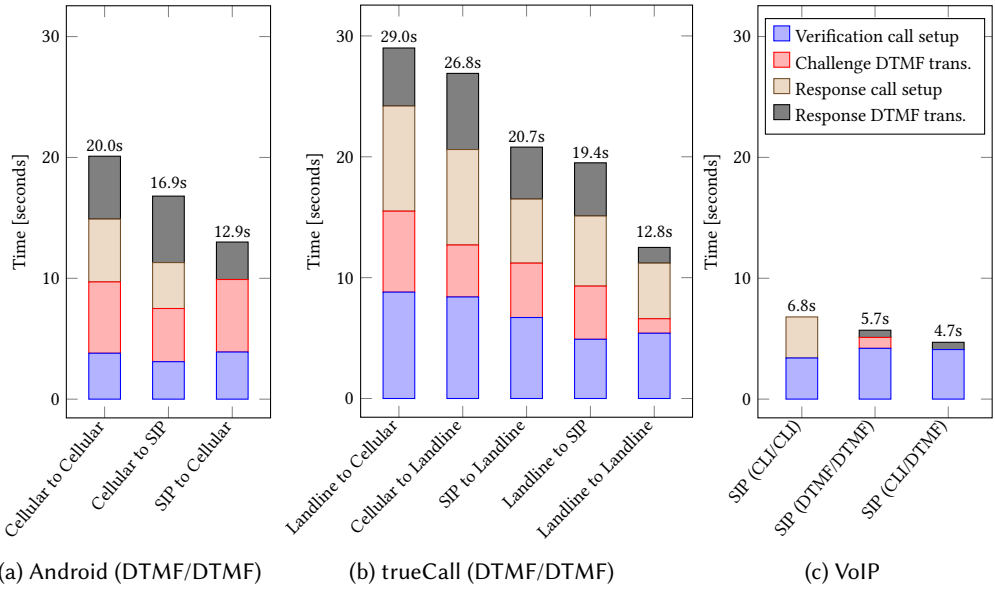

(a) Android (DTMF/DTMF)          (b) trueCall (DTMF/DTMF)          (c) VoIP

Fig. 10. Cost breakdowns in CIV

DTMF challenge (since the challenge is embedded in CLI; see Figure 10 (c)). Among these prototypes, 'CLI/CLI' incurs the longest delay because sending the response through CLI requires a call setup, which involves a significant cost. The other two prototypes use an *existing* initial call to send the DTMF response and hence are free from this call setup cost.

## 5.2 Other platforms

Prototypes built on trueCall and Android are limited by various platform constraints (see Case 3 in Figure 6). These constraints do not prevent the proof-of-concept implementation of CIV, but they increase the latency.

**Android**. Our CIV implementation on Android phones is network-agnostic as it only requires the DTMF transmission which is supported in all cellular networks. We performed experiments in GSM networks to set a benchmark of baseline performance. As shown in Figure 9, when a cellular phone calls a cellular phone, the total latency of CIV is about 20 seconds. This delay is due to two main factors. First, the third-party API that we use to build a CIV-enabled phone app does not support sending DTMF during a call. We overcome this by appending DTMF digits as an extension of a phone number upon dialing, but this requires two call setups (for sending the challenge and response respectively). Second, as the third-party API does not support automatically recognizing DTMF, we had to overcome this by playing out the audio sound of the DTMF tones via a speaker and decoding them into digits. It is worth noting that when a SIP phone calls a cellular phone, our proof-of-concept implementation is not affected by the first factor; the caller (SIP) is able to send back the DTMF response through the initial call because the SIP platform supports sending DTMF during the call. This removes the cost of "response call setup" (see Figure 10 (a) for a breakdown), but the overall delay of 'SIP to Cellular' is still dominated by the slow recognition of DTMF tones.

**trueCall**. As shown in Figure 9, when a landline phone calls a landline phone, the total delay of CIV is about 13 sec. The trueCall supports transmitting in-band DTMF and the automatic recognition of DTMF during a call. But the performance is limited by the lack of the *call waiting* function in the trueCall hardware. As a result, we need a full call setup to send the response, which is a significant cost component (see Figure 10 (b)). This affects all experiments that involve a landline phone with a trueCall box. Also, since the landline phone is connected to a PSTN network via an analogue line, transmitting DTMF between the landline phone and other networks incurs more delays. The worst case is when CIV is run between a landline phone and an Android phone; the total latency is 29 sec. This is because the hardware limitation of trueCall is compounded by the slow recognition of DTMF tones in our Android prototype. In Section 6, we will discuss how the performance can be substantially improved once the underlying platform constraints are removed.

## 5.3 Accuracy of DTMF transmission

Successful authentication in CIV depends on the reliable transmission of the response through DTMF. We discuss the accuracy of DTMF transmission in different conditions below.

**Out-of-band DTMF**. When transmitted out-of-band (see Figure 1), DTMF signals are embedded in the control messages as digital data. The transmission of digital data in the control channels of telecommunication systems is highly reliable by design since that is crucial for the basic functioning of the telecommunication system.

**In-band DTMF**. When transmitted in-band, DTMF signals are sent through the voice channel as part of an audio stream. The accuracy of the DTMF decoding depends on two parameters: the duration of each DTMF tone ('mark') and the gap between the two tones ('space'). In a SIP system, the minimum values for the mark and space are defined to be 40 ms according to RFC 4733. In our SIP prototypes, we set both parameters to be 50 ms. Under this setting, we observe that the decoding of DTMF in our experiments is 100% accurate. However, in a PSTN system where DTMF tones are sent as analogue data, the DTMF transmission is more susceptible to noise. To determine a suitable setting for sending DTMF reliably in analogue phone lines, we perform an experiment using two TrueCall boxes, whereby we specify different mark/space values for sending 4-digit DTMF signals and evaluate the rate of successful decoding.

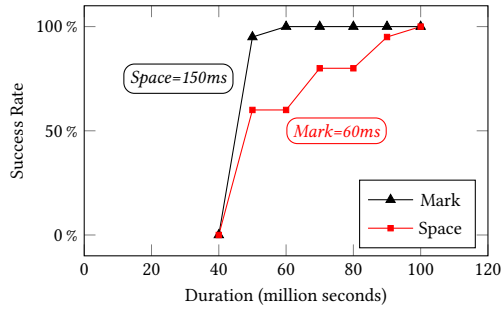

Fig. 11. Success rate of sending 4-digit DTMF code through analogue lines with different mark/space values

Figure 11 summarizes the success rate after we choose a random 4-digit code for DTMF transmission and repeat the experiment 20 times for each mark/space setting. First, we set the space to a generous value of 150 ms and examine the effect of varying the mark value. We observe that the success rate reaches 100% when the mark value equals 60 ms (or above). Next, we set the mark value to 60 ms and vary the space value. We observe that the success rate reaches 100% when the space value equals 100 ms (or above). Based on these experiments, for trueCall-based CIV prototypes, we choose mark = 100 ms and space = 100 ms. This ensures reliable transmission of the response through analogue DTMF signals in our landline-phone prototypes.

## 5.4 Lengths of the challenge and response

In the design of CIV, we choose $n = 4$ for the number of digits in the challenge, as a reasonable trade-off between security and performance. Figure 12 shows the variation of the DTMF transmission time for different values of $n$. As expected, in all cases, the transmission time increases linearly with $n$. We would like to draw attention to the 'SIP to SIP' measurements since they involve only the transmission of digital DTMF values (not analogue tones). This is the trend as analogue phone lines are being phased out. As shown in the figure, the 'SIP to SIP' measurements have a relatively flat slope; increasing $n$ hardly increases the delay. This shows the flexible scope of choosing a bigger $n$ with little performance degradation.

## 6 DISCUSSION

In this section, we discuss various aspects of the CIV system, including security, deployment, and limitations.

### 6.1 Downdegrading attack

In the CIV design, the caller needs to indicate support for CIV in the initial call. This can be achieved by appending a special CIV flag in the caller name. When the caller name for a phone number is registered in a CNAM database, the CIV flag is also saved there. Hence, when a terminating network retrieves the caller name from CNAM databases for an incoming call, it recognizes that the caller supports CIV. Alternatively, the caller name may be sent in the call along with the caller ID.

This design supports callers who *actively* want to be verified, e.g., to increase the likelihood that the called party will answer the call. However, an attacker may launch a downgrading attack to bypass CIV: spoofing a caller ID and transmitting a false caller name without any CIV flag. If the caller ID/name is registered in CNAM databases, the terminating network can look up the databases, and use the retrieved caller name to overwrite the received caller name. However, CNAM databases are not used everywhere. Also, the terminating network might simply use the received caller name instead of looking up CNAM databases (the latter involves paying a 'dip' fee). Hence, it is possible that even though a caller ID is registered as CIV-enabled, an attacker may spoof that caller ID without invoking the CIV process.

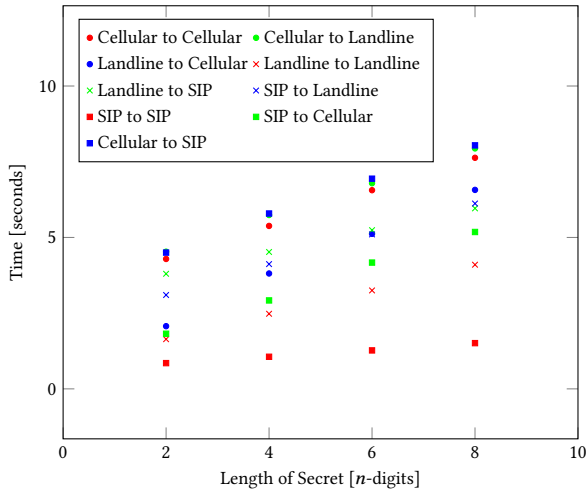

Fig. 12. DTMF transmission time versus number of digits

But the above downgrading attack has limited effects. In the worst case, the called party's phone rings with a display of a caller ID, but the CIV program shows a warning "caller not verified". This is actually an intended outcome for the CIV design. A successful attack should involve spoofing a number that the attacker does not own and showing "caller verified" on the callee's phone. This is unlikely as we explain below. We note that it is possible for a user to save a phone number as a local contact on the phone with a CIV flag. This ensures that the CIV program is always invoked when receiving a call with a display of that number.

## 6.2 Attacking the challenge-response protocol

The main idea of CIV lies in how we define "authentication": based on if the caller can prove the possession of a phone number by answering a challenge sent to that number. In practice, when a user registers a phone number with a system, the system checks if the provided number is legitimate by calling the number or sending an SMS code. In our design, we automate the verification process by sending a random 4-digit code as part of the CLI (preferred) or through DTMF. In our threat model (Section 3.1), we assume that a spoofing attacker has no control over intercepting calls in telecommunication systems. Hence, if they do not own the phone number, they cannot receive the challenge. This leaves them with the only option of guessing the challenge and forging a response. With the 4-digit code in the challenge, the chance of forging a valid response is $10^{-4} = 0.01\%$ (which can be further reduced by increasing the number of digits). This does not completely eliminate the theoretical attack, but it massively reduces the success probability, hence serving as a deterrent. We note that CIV *always* connects a call regardless of the verification result. Hence, repeated failures of guessing will alert the user.

## 6.3 Denial-of-service (DoS)

In CIV, a malicious caller may spoof an arbitrary caller ID when making a call. The goal is not to pass the CIV verification but to leverage the callee to call back to the spoofed number, e.g., to cause a nuisance. We address this threat in three ways. First, CIV is invoked only when the caller indicates support for CIV in the initial call. A user can further configure CIV to perform verification only for certain numbers (e.g., domestic, non-premium numbers). Second, we propose using 'CLI/DTMF' as our recommended implementation of sending challenge/response (see Figure 7). Since the callee uses a modified CLI to transmit the 4-digit challenge in an *abandoned* call, there is no cost to the callee[2]. If the verification call reaches an unsuspecting phone user who has no CIV installed, it will be shown as a silent *missed call* with a non-dialable four-digit number. If the user's phone has CIV installed, CIV can recognize unsolicited verification calls and automatically filter them out. We note that the attacker can always make silent *missed calls* to a target user directly, but the attack is deterred since the attacker can be traced by checking the call detail record (CDR) at the telephone exchanges. In CIV, the attacker may leverage the callee to make a reflected DoS attack against a user but the attack can still be traced by correlating the two related CDRs. Any such DoS attack actually has a limited effect: in the worst case, it leaves a silent *missed* call with a non-diable number, but the call can be easily filtered by the user's phone (say by installing a CIV program). Third, as a long-term solution, we propose integrating CIV into the telecom cloud so the challenge-response protocol is performed between the switches of the two communicating carriers in the telecom cloud rather than the two phones. This will be the best way to deploy CIV as we elaborate below.

## 6.4    Stages of deployment and optimization

For calls between landline, cellular and VoIP phones, we have explored the most practical way to implement CIV for each call scenario within constraints; all this is done without cooperation from network providers. A unified and optimal implementation of CIV is possible by integrating CIV into the networks in a three-stage deployment.

**Stage 1 (short term).** This stage involves proof-of-concept demonstrations of CIV on the users' phones as we have done in this work. Without any cooperation from network providers, we show CIV can be implemented for all three different telecom networks (PSTN, cellular and VoIP) without modifying the existing infrastructure. This presents probably the best that can be done at the user's end under various platform constraints. Some of the constraints can be easily removed, e.g., by adding call-waiting and DTMF functions on some phones. Other constraints are more fundamental, e.g., access to the CLI modification function. Understanding these constraints lays the foundation for the next stage of CIV deployment.

**Stage 2 (medium term).** This stage involves integrating CIV into the *terminating network*. This removes one of the most important constraints in implementing the 'CLI/DTMF' method as described in Figure 7: namely, access to the facility of modifying CLI for sending the challenge. We note that the integration of CIV can be done *independently* by any terminating network without cooperation from other providers. When receiving a call with the CIV flag, the terminating network performs the verification process on behalf of its subscriber and relays the caller ID as well as the verification result to the subscriber's phone.

**Stage 3 (long term).** A network provider can extend their service in Stage 2 to perform CIV on behalf of the caller as well. When more networks implement CIV and support both the caller and the callee, it may reach a point where networks commonly support CIV and there is no need to install CIV on the user's device anymore. CIV then essentially becomes a new service in the telecom cloud (as an enhanced version of CLI with verified caller ID), which users can subscribe to.

---

[2]Incurring no cost to the callee was highlighted as important during our meetings with telecom providers (particularly, Squire Technologies and REDtone Telecommunications) to gather design requirements.

Table 1. Comparison with related works

| Scheme | Auth subject | Delay | Deterministic authentication | Heuristic authentication | Requires registration | Requires TTP | Diff. legitimate spoofing | Works with SIP | Works with SS7 | SIP phone prototype | Mobile phone prototype | Landline phone prototype | Hetero. networks experiment |
|---|---|---|---|---|---|---|---|---|---|---|---|---|---|
| **Top-down** | | | | | | | | | | | | | |
| STIR/SHAKEN [8] | Carrier | N/A | ●(b) | | ●(r) | ●(r) | | ●(b) | ◐(b) | | ●(b) | | ●(b) |
| AuthLoop [16] | Call center | 4.8–8.9 s | ●(b) | | ●(r) | ●(r) | | ●(b) | ●(b) | | ●(b) | | |
| AuthentiCall [15] | Caller ID | 1–1.41 s | ●(b) | | ●(r) | ●(r) | | ●(b) | ●(b) | | | ◐(b) | |
| **Bottom-up** | | | | | | | | | | | | | |
| CallerDec [13] | Caller ID | 8.4–20 s | | ●(r) | | | | | ◐(b) | | ●(b) | | ●(b) |
| CEIVE [4] | Caller ID | 10–23 s | | ●(r) | | | | ●(b) | | | ●(b) | | |
| CIV (this paper) | Caller ID | 4.7–29 s | | ◐(r) | | | ●(b) | ●(b) | ●(b) | ●(b) | ●(b) | ●(b) | ●(b) |

●/◐: have a full/partial *undesirable* property (red). ●/◐: have a full/partial *desirable* property (blue).

The challenge-response process is done between the switches of the two communicating carriers within the telecom cloud.

**Example.** trueCall [24] has already been integrated into the UK telecom network so it is available as a telephone service that users can subscribe to. Users do not need to install any trueCall box at their end, since the service is virtually available in the cloud. This is realized by adding a software "hook" in the cloud to handle calls on behalf of the subscribed trueCall users. We expect that CIV will follow a similar approach for the integration into telecom clouds.

## 6.5 Limitations

Our current work on CIV has several limitations. First, the verification delay for landline and cellular phones is high (12−29 seconds). This is mainly because we do not have access to the facility to modify CLI for the landline and cellular phones (only the switches in the network can modify CLI). Hence, we are not able to use the most efficient 'CLI/DTMF' method to implement the challenge-response process. Second, our SIP prototype based on 'CLI/DTMF' reports 4.7 seconds delay, which is closer to being practical, but there is still room for improvement. As shown in Figure 10 (c), this delay is dominated by the 'verification call setup', which involves not only routing the call but also, more importantly, allocating resources along the call path to prepare for the ensuing telephone conversation when the call is answered. However, in our case, the verification call is only to transmit a short challenge, not intending for a conversation. Hence, the call setup using `INVITE` takes longer than necessary. This delay can be substantially reduced by using a different signaling mechanism (e.g., out-of-band `INFO` or `NOTIFY` messages), but this needs to be done between SIP providers rather than from SIP phones. Third, the current implementation of CIV for SIP adds an extra flow of `INVITE` signaling for the verification call, which may add a burden on some networks and affect the termination success rate. As with the previous one, this can be addressed by using a different (out-of-band) signaling method between SIP providers to transmit the challenge. It should be possible to overcome all these limitations by integrating CIV into the telecom cloud (Stage 3 deployment), which we plan to investigate in further research.

## 7  RELATED WORK

We broadly divide previous solutions into two types: top-down and bottom-up. A top-down solution involves introducing a trusted third party (TTP), while a bottom-up solution does not need one. Table 1 presents an overview of the representative schemes for each type. There are several *undesirable* properties that we wish to avoid. In particular, we want to avoid relying on a TTP[3] and an extra registration process if possible. We consider two types of authentication: heuristic and deterministic. The former performs authentication *probabilistically* based on heuristics (e.g., network characteristics and prior training data), but the authentication result may vary under different conditions and troubleshooting failures can prove difficult. The latter performs authentication *deterministically* based on explicit rules (e.g., possession of a secret key or a secret number), which is considered more desirable. In addition, it is necessary that a caller ID authentication solution should distinguish legitimate and illegitimate modifications of a caller ID. Finally, the solution should work with heterogeneous networks (SS7/SIP) and be tested in such conditions.

### 7.1  Top-down

STIR/SHAKEN was jointly developed by several telecom companies. To our knowledge, there is no peer-reviewed academic paper on the original proposal of STIR/SHAKEN, but the scheme is described in a series of IETF RFCs [8]. STIR/SHAKEN critically relies on trusted certificate authorities (CAs) in a PKI to issue digital certificates for the subscribed carriers. It authenticates the carrier (not the caller) based on their possession of a unique private signing key. It is designed to work with an IP-based SIP network and has been commercially deployed on VoIP and (IP-based) mobile phones. STIR/SHAKEN does not authenticate any caller ID, or distinguish legitimate/illegitimate spoofing; it relies on the carrier's 'word of mouth' for attesting to the authenticity of the caller ID.

AuthLoop was proposed by Reaves et al. at USENIX Security'16 [16] based on adapting TLS 1.2 to a telephony network. Same as STIR/SHAKEN, it requires a PKI (called a "Telephony PKI" in their paper). This scheme is designed for a client-server setting, where the server (a call center) is authenticated based on digital signatures, but the client caller is not authenticated. The system does not work with existing SIP or SS7 signaling. A prototype of AuthLoop was implemented between PCs (as clients and servers) but not on telephones or connected to telephone networks. The verification delay was reported to be 4.8–8.9 seconds.

AuthentiCall was proposed by Reaves et al. at USENIX Security'17 [15]. To address the lack of caller authentication in AuthLoop, the authors propose to introduce a trusted central server. Telephone users need to register their numbers (including legitimately modified numbers) with this server in an enrollment process to obtain certificates. When the user makes a call, the phone first contacts the central server, and then the server mediates the authentication process between the caller and the callee. A prototype was implemented using an Android app connected to a server through the Internet, but the app was not connected to telephone networks. The authors report 1–1.41 seconds delay for the verification process (and 22–25 seconds for the enrollment process, which is not included in Table 1).

### 7.2  Bottom-up

CallerDec was proposed by Mustafa et al. at DSN'14 [12] (with a journal version in 2016 [13]). It is designed for a circuit-switched telephone network. It authenticates caller ID by calling back the number. This scheme is the closest to ours. It has the advantage of not needing any PKI or extra registration. However, CallerDec authenticates the caller ID *probabilistically* based on applying heuristics to infer the caller's state. The authentication outcome critically depends on physical network characteristics, such as timing in the call setup, as well as the choice of classifiers and the prior training data. CallerDec, on its own, does not distinguish legitimate and illegitimate spoofing; instead, it requires the user to press keys on the keypad to indicate if a spoofed call is legitimate or not. A prototype based on an Android mobile phone was implemented and tested in a circuit-switched cellular (3G) network. The verification delay was reported to be 8.4 (call-based) and 20 (SMS-based) seconds.

---

[3]Recall a TTP is "a third party that can break your security policy" [1].

|  | Set up cost | | | | Ongoing cost |
|---|---|---|---|---|---|
|  | Phone update | Telco update | Cert reg. | Key man. | Cert main/renew |
| STIR/SHAKEN | ✓ | ✓ | ✓ | ✓ | ✓ |
| CIV | Optional | ✓ | – | – | – |

Table 2. Comparison of cost between STIR/SHAKEN and CIV (stage 3)

CEIVE was proposed by Deng et al. at MobiCom'18 to authenticate caller ID in a 4G wireless network [4]. It uses a similar call-back idea as CallerDec to infer the caller's state. The main difference is that it is a *callee-only* solution without needing any cooperation from the caller. Same as CallerDec, it authenticates the caller ID *probabilistically* based on heuristics. The authors acknowledge that this is reliable for a single carrier, but the performance varies for a different carrier or across carriers. CEIVE does not distinguish legitimate and illegitimate spoofing. It was implemented on an Android phone and tested on a 4G network. A verification delay of 10−23 seconds was reported.

### 7.3 Comparison between STIR/SHAKEN and CIV on cost/benefit

We perform a cost/benefit comparison between STIR/SHAKEN and CIV. Here, we only consider the Stage 3 deployment of CIV (Section 6.4) as it is the most practical for a real-world implementation.

**Cost**. We consider two types of costs: one-time setup cost and ongoing cost. Table 2 summarizes the comparison result. For the stage 3 deployment of CIV, the challenge-response process is performed between the switches of the two communicating carriers in the telecom cloud. This should only require a software update in the telecom cloud. The software update on the user's phone is optional. We may update the phone display showing the verification outcome of CIV, but it is also possible to inform the user via pe-recorded audio (i.e., whether the caller ID is verified or not) when she answers the phone. The latter can work with existing phones without modification as we have done with the trueCall prototype. By comparison, different levels of attestation (A, B, C) in STIR/SHAKEN cannot be communicated succinctly via audio [18]; a STIR/SHAKEN compatible phone is required to display this complex information. The more significant cost in STIR/SHAKEN involves certificate registration and the private key management by each carrier. We note that it is unsafe to manage private keys in software. The best practice requires private keys to be managed in hardware security modules (HSMs) [1], which are costly. Furthermore, each carrier must pay ongoing costs to CAs for certificate maintenance (e.g., possible revocation and replacement) and annual renewal. These certificate-related costs do not occur in CIV.

**Benefit**. The primary goal of STIR/SHAKEN and CIV is to prevent caller ID spoofing attacks by scammers. Hence, we evaluate the benefit in terms of how effectively the solution can achieve its goal. We focus on analyzing the main mode of operation by scammers: they call from an overseas VoIP provider but use a spoofed *local* number which they do not own in order to deceive the receiver. The digital signature in STIR/SHAKEN does not address this problem per se. A C-level attestation only states that the call is routed via an international gateway but says nothing about the authenticity of the caller ID [18]. (Usually, the gateway lacks the knowledge to tell whether an overseas caller is authorized to use a local number or not.) CIV addresses this problem by sending a verification call to the local number, hence preventing spoofing attacks without requiring the cooperation of the overseas VoIP provider or the gateway. This shows that, to address caller ID spoofing as an international problem, it is actually sufficient to deploy CIV domestically.

## 8   CONCLUSION

We propose CIV, a new solution to authenticate caller ID in heterogeneous telephone networks without a public key infrastructure. CIV authenticates the caller ID through a challenge-response protocol; it distinguishes legitimate and illegitimate spoofing based on whether the caller owns the phone number; it supports both SS7 and SIP; and it has been implemented on all three types of phone systems and tested across heterogeneous networks to demonstrate feasibility. Contrary to the common belief by the FCC and regulators in some other countries that STIR/SHAKEN is the only solution, our work shows that alternatives exist and that they can be far more cost-effective than STIR/SHAKEN. We hope this will encourage more research into bottom-up solutions to address caller ID spoofing without relying on any trusted third party or a PKI.

## ACKNOWLEDGEMENT

This work is supported by EPSRC (EP/T014784/1). We want to thank anonymous reviewers for their invaluable feedback and the following people for many helpful comments: Adrian von Mühlenen of the University of Warwick; Eric Priezkalns of Risk & Assurance Group; Sanjeev Verma, Basi Thomas, Ben Teversham and Michael Karakashian of Squire Technologies; David Maxwell, Travis Russell and Kwok Onn Looi of GSMA; Yongdae Kim of Korea Advanced Institute of Science and Technology; Siamak Shahandashti of the University of York; Syed Ahsun Abbas Rizvi and Khurram Mushtaq of REDtone Telecommunications Pakistan.

## REFERENCES

[1]   Ross Anderson. 2020. *Security engineering: a guide to building dependable distributed systems.* John Wiley & Sons.
[2]   Charles Beumier and Thibault Debatty. 2022. Attack detection in ss7. In *International Conference on Multimedia Communications, Services and Security.* Springer, 11–20.
[3]   Stanley T Chow, Christophe Gustave, and Dmitri Vinokurov. 2009. Authenticating displayed names in telephony. *Bell Labs Technical Journal,* 14, 1, 267–282.
[4]   Haotian Deng, Weicheng Wang, and Chunyi Peng. 2018. Ceive: combating caller id spoofing on 4g mobile phones via callee-only inference and verification. In *24th Annual International Conference on Mobile Computing and Networking,* 369–384.
[5]   Lee Dryburgh and Jeff Hewett. 2005. *Signaling System No. 7 (SS7/C7): protocol, architecture, and services.* Cisco press.
[6]   FCC. 2023. Caller id spoofing. https://www.fcc.gov/spoofing. (2023).
[7]   FCC. 2023. Combating spoofed robocalls with caller id authentication. https://www.fcc.gov/call-authentication. (2023).
[8]   IETF. 2023. Secure Telephone Identity Revisited (STIR). https://datatracker.ietf.org/wg/stir/documents/. (2023).
[9]   Michael Lesk. 2014. Caller id: whose privacy? *IEEE security & privacy,* 12, 2, 77–79.
[10]  Jikai Li, Fernando Faria, Jinsong Chen, and Daan Liang. 2017. A mechanism to authenticate caller id. In *World Conference on Information Systems and Technologies.* Springer, 745–753.
[11]  Ahmadreza Montazerolghaem. 2022. Softwarization and virtualization of voip networks. *The Journal of Supercomputing,* 1–33.
[12]  Hossen Mustafa, Wenyuan Xu, Ahmad Reza Sadeghi, and Steffen Schulz. 2014. You can call but you can't hide: detecting caller id spoofing attacks. In *44th Annual IEEE/IFIP International Conference on Dependable Systems and Networks.* IEEE, 168–179.
[13]  Hossen Mustafa, Wenyuan Xu, Ahmad-Reza Sadeghi, and Steffen Schulz. 2016. End-to-end detection of caller id spoofing attacks. *IEEE Transactions on Dependable and Secure Computing,* 15, 3, 423–436.
[14]  Ofcom. 2022. Number spoofing scams. https://www.ofcom.org.uk/phones-telecoms-and-internet/advice-for-consumers/scams/phone-spoof-scam. (2022).

[15] Bradley Reaves, Logan Blue, Hadi Abdullah, Luis Vargas, Patrick Traynor, and Thomas Shrimpton. 2017. Authenticall: efficient identity and content authentication for phone calls. In *26th USENIX Security Symposium*, 575–592.

[16] Bradley Reaves, Logan Blue, and Patrick Traynor. 2016. Authloop: end-to-end cryptographic authentication for telephony over voice channels. In *25th USENIX Security Symposium*, 963–978.

[17] Merve Sahin, Aurélien Francillon, Payas Gupta, and Mustaque Ahamad. 2017. Sok: fraud in telephony networks. In *2017 IEEE European Symposium on Security and Privacy (EuroS&P)*. IEEE, 235–250.

[18] Imani Sherman, Daniel A Delgado, Juan E Gilbert, Jaime Ruiz, and Patrick Traynor. 2021. Characterizing user comprehension in the stir/shaken anti-robocall standard. In *49th Research Conference on Communication, Information and Internet Policy*.

[19] 2021. The UK's PSTN network will switch off in 2025. (Oct. 11, 2021). https://business.bt.com/why-choose-bt/insight s/digital-transformation/uk-pstn-switch-off/.

[20] TransNexus. 2022. Reply comments on shaken extensions and effectiveness. https://transnexus.com/blog/2022/shake n-effectiveness-extensions-reply-comments/. (2022).

[21] TransNexus. 2022. Robocalls up sharply in october. https://transnexus.com/blog/2022/robocalls-up-sharply-october/. (2022).

[22] TransNexus. 2021. Service provider sti fee changes for 2021. https://transnexus.com/blog/2021/sti-provider-rate-cha nges/. (2021).

[23] TransNexus. 2022. Stir/shaken statistics from october 2022. https://transnexus.com/blog/2022/shaken-statistics-octo ber/. (2022).

[24] trueCall. 2023. https://www.truecall.co.uk/. (2023).

[25] Truecaller. 2023. https://www.truecaller.com/. (2023).

[26] Truecaller. 2022. 2022 U.S. Spam & Scam Report. https://truecaller.blog/2022/05/24/truecaller-insights-2022-us-spam -scam-report/. (2022).

[27] Huahong Tu, Adam Doupé, Ziming Zhao, and Gail-Joon Ahn. 2016. Sok: everyone hates robocalls: a survey of techniques against telephone spam. In *2016 IEEE Symposium on Security and Privacy (SP)*, 320–338. DOI: 10.1109/SP.2 016.27.

[28] Huahong Tu, Adam Doupé, Ziming Zhao, and Gail-Joon Ahn. 2017. Toward standardization of authenticated caller id transmission. *IEEE Communications Standards Magazine*, 1, 3, 30–36.

[29] John G Van Bosse and Fabrizio U Devetak. 2006. *Signaling in telecommunication networks*. Vol. 87. John Wiley & Sons.

[30] Xinyuan Wang and Ruishan Zhang. 2011. Voip security: vulnerabilities, exploits, and defenses. In *Advances in Computers*. Vol. 81. Elsevier, 1–49.

[31] YouGov. 2019. Don't call me: nearly 90% of customers won't answer the phone anymore. https://martech.org/dont-c all-me-nearly-90-of-customers-wont-answer-the-phone-anymore-study/. (2019).

[32] James Yu. 2021. An analysis of applying stir/shaken to prevent robocalls. In *Advances in Security, Networks, and Internet of Things*. Springer, 277–290.